\documentclass{svproc}

\usepackage{url}
\usepackage{graphics}
\usepackage{graphicx}
\usepackage{multicol}
\usepackage{footmisc}
\usepackage{algorithm}
\usepackage{flushend}
\usepackage{algpseudocode}
\usepackage{booktabs}
\usepackage{makecell}
\usepackage{eqnarray}
\usepackage{amsmath}
\usepackage{amssymb}
\usepackage{color}
\usepackage{cite}
\usepackage[colorlinks, citecolor=blue,linkcolor=blue,anchorcolor=blue,citecolor=blue]{hyperref}

\begin{document}

\mainmatter

\title{A Three-Dimensional Pursuit-Evasion Game Based on Fuzzy Actor-Critic Learning Algorithm}

\titlerunning{ }  

\author{Penglin Hu}

\authorrunning{ } 

\institute{Northwestern Polytechnical University, Xi'an Shaanxi 710129, China\\
\email{penglinhu@mail.nwpu.edu.cn}}

\maketitle 
\begin{abstract}
Most of the existing research on pursuit-evasion game (PEG) is conducted in a two-dimensional (2D) environment. In this paper, we investigate the PEG in a 3D space. We extend the Apollonius circle (AC) to the 3D space and introduce its detailed analytical form. To enhance the capture efficiency, we derive the optimal motion space for both the pursuer and the evader. To address the issue arising from a discrete state space, we design a fuzzy actor-critic learning (FACL) algorithm to obtain the agents' strategies. To improve learning performance, we devise a reward function for the agents, which enables obstacle avoidance functionality. The effectiveness of the proposed algorithm is validated through simulation experiments.
\end{abstract}

\section{Introduction}
In recent years, as a typical adversarial decision-making problem, the pursuit-evasion game (PEG) has attracted widespread attention \cite{Weintraub}. In 1965, the PEG, which consists of a pursuer and an evader, was proposed \cite{Ho-Y}. The PEG has been widely applied in military and civil fields, such as missile defense and interception \cite{Lee}, air combat \cite{Biediger}, search and rescue \cite{Zhang}, transportation management \cite{Yan-R}, and so on. However, traditional PEG research is mostly limited to 2D space, which is inadequate when dealing with realistic scenarios involving complex 3D  environment in the real world.

In \cite{Awheda}, the location of the player is determined using the Apollonius circle (AC) to define the capture area. Then, a formation control method is employed to construct a reward function for the learning tracker, which is used to adjust the fuzzy logic control of the learning tracker. Researchers have developed a multiple pursuer multiple evader PEG model based on the AC method. They determine the optimal control strategy through the angular and distance between the players \cite{Garcia}. Based on the AC, the player's dominant region is divided. Meanwhile, a sacrifice strategy is proposed, which means that when both evaders are likely to be captured, one evader sacrifices itself to help the other one \cite{Yan2021}. By utilizing the Voronoi diagram to partition the area and define the dominant region of the player, a relay tracing scheme based on the regional partition is proposed \cite{PanT2022}. In \cite{Selvakumar2019}, through the division of the state space, a risk measurement function is designed, and a nonlinear state feedback control strategy is developed for the evader, enabling the implementation of a PEG with terminal constraints.

Compared to the 2D environment, the PEG exhibits higher complexity and poses greater challenges in 3D environments. In 3D space, the movement trajectories of agents are more variable, and the impact of obstacles and terrain is more significant. Therefore, how to design effective algorithms to solve the 3D PEG has become a hot and difficult issue in current research. In \cite{Kartal2021}, the authors utilize Pontryagin's minimum principle to derive the optimal actions for the players. By representing the agents' attitudes using unit quaternions, they are able to solve the 3D PEG. Furthermore, authors have applied the fuzzy actor-critic learning (FACL) method to the PEG problem involving unmanned aerial vehicles (UAVs) in continuous spaces\cite{Al-Mahbashi2020}. Xie et al. studied the PEG in a 3D environment and proposed the concept of occupied angle to represent the pursuer's dominant region, designing the control strategies for both the pursuer and the evader \cite{X-Fang2020}. 

Based on the above background, this paper proposes a 3D PEG algorithm based on the FACL \cite{Schwartz} algorithm. The contribution of this paper can be summarized as follows. First, we extend 2D AC to its generalized form in 3D space and establish a detailed analytical form to design the optimal movement areas for both players of the game. In \cite{Awheda,Garcia,Yan2021,PanT2022}, AC is utilized to address the 2D PEG problem. In \cite{Kartal2021} and \cite{Al-Mahbashi2020}, both studies employed the 3D Euclidean distance to design PEG strategies. Although the concept of occupied angle is adopted in \cite{X-Fang2020}, its application scenarios are limited. Therefore, the generalized AC proposed in this paper has good advantages in solving 3D PEG. Second, we design the FACL algorithm for agent training and provide the system's update process equation. Finally, we design an efficient and reasonable reward function based on the concept of artificial potential field. 

The remainder of this paper is organized as follows. In Section 2, a mathematical description of the PEG problem is presented. Section 3 introduces the proposed algorithm. Section 4 shows the simulation results. Section 5 summarizes the paper and provides future research directions.

\section{Problem Formulation}
\begin{figure}[!ht]
	\centering
	\includegraphics[width=0.9\linewidth]{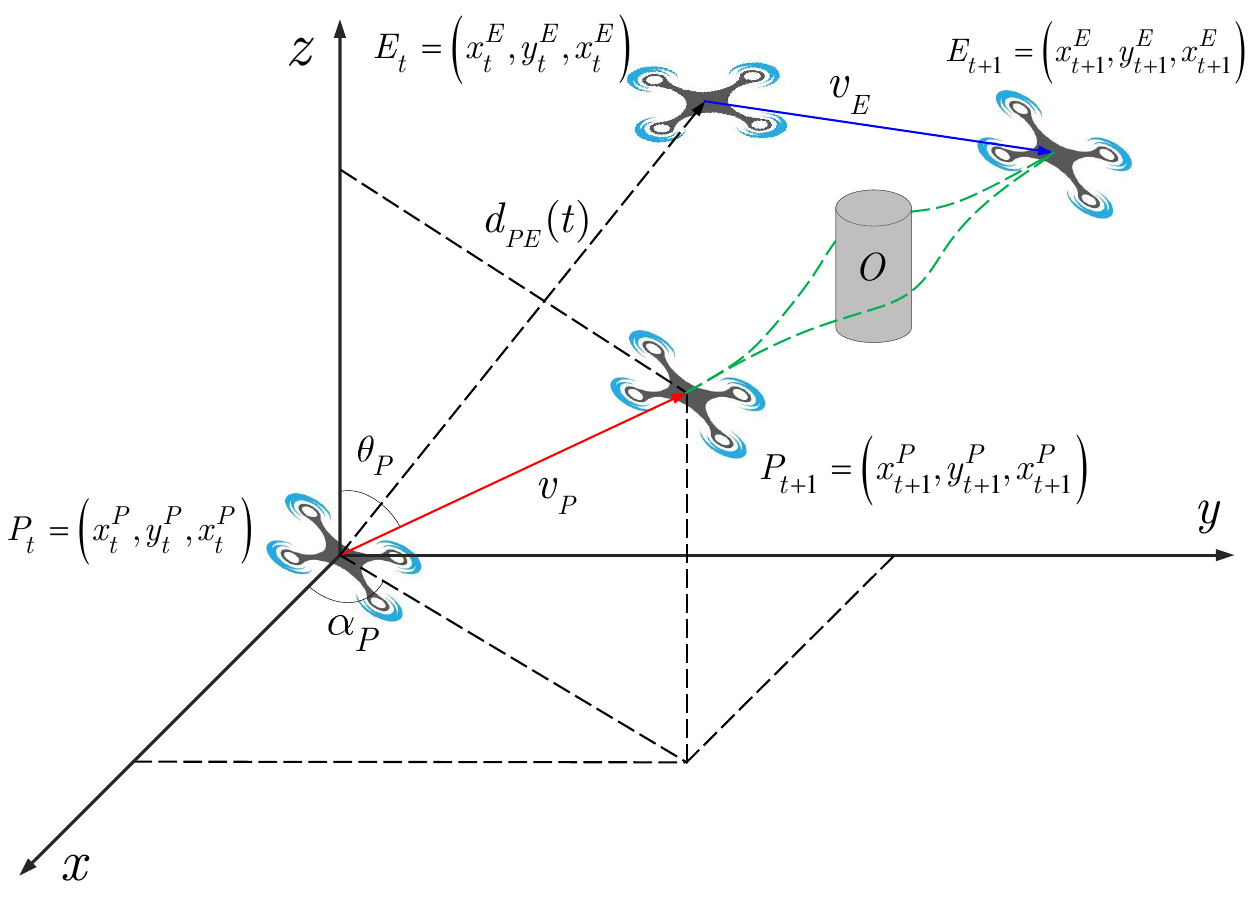}
	\caption{Illustration of the PEG model, where $P_t$ represents the pursuer, $E_t$ represents the evader, and the cylinder represents the obstacle $O$}
	\label{PE-model}
\end{figure}

The pursuer generates its trajectory prediction by assessing factors such as the evader's position and velocity. As shown in Fig.~\ref{PE-model}, in 3D space, the agent's kinematic model is described as \eqref{kinematic equation of the agent} 
\begin{equation}
\label{kinematic equation of the agent}
\begin{aligned}
{x_{t + 1}} &= {x_t} + v\sin \theta \cos \alpha \\
{y_{t + 1}} &= {y_t} + v\sin \theta \sin \alpha \\
{z_{t + 1}} &= {z_t} + v\cos \theta,
\end{aligned}
\end{equation}
where $(x_t,y_t,z_t)$ is the agent’s position, $v$ is the agent’s speed, $\alpha$ is the angle between the projection of speed $v$ onto the $x$-$y$ plane and the $x$-axis, and $\theta$ is the angle between speed $v$ and the $z$-axis. $U= {[\alpha ,\theta ]^ \top }$ is the steering angle of the agent, and also the output of the controller. We assume that agents are aware of each other’s location information, but they are not aware of the strategy adopted by each other. To ensure that the agent's motion adheres to real-world constraints, the agent’s steering angle is limited in the interval of $\left[ {-\frac{\pi }{4},\frac{\pi }{4}} \right]$.

In the PEG, players optimize their respective strategies through continuous trial and error and reflection, and gradually improve through interactions with the environment. The pursuer's goal is to capture the evader in the shortest time possible, while the evader's goal is to evade capture or delay the time of being captured. We define the terminal condition for the PEG as follows: the PEG process ends at time $ {t_f} $ if the distance between the pursuer $P$ and the evader $ E $ satisfies $ {d_{PE}}({t_f}) \le {d_{s}} $, where $ {d_{s}} $ represents the capture distance. Additionally, if the time exceeds the designated threshold, the PEG is over.

\begin{figure}[!ht]
\centering
\includegraphics[width=1\columnwidth]{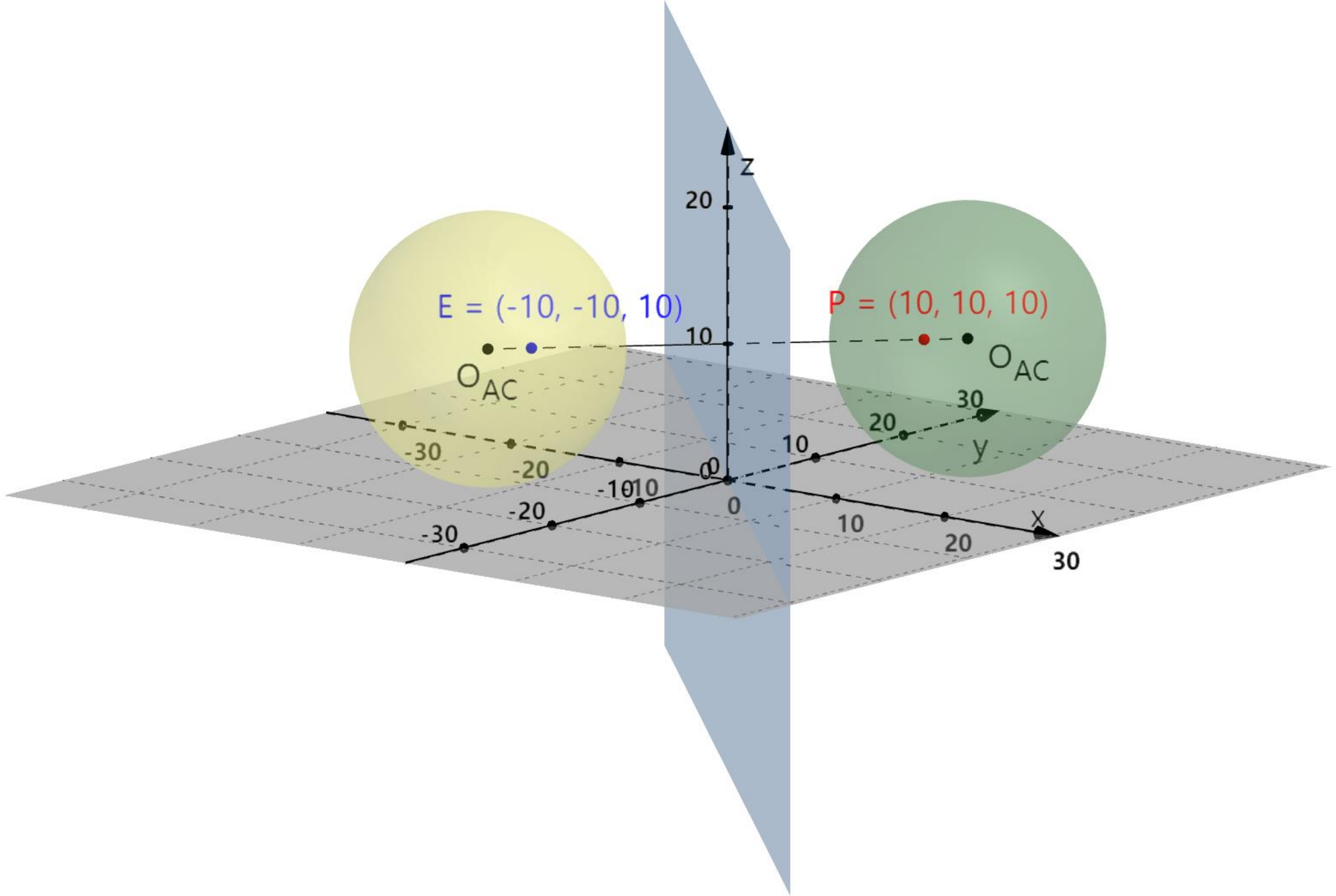}
\caption{Illustration of the generalized AC, where $P=(10,10,10)$, $E=(-10,-10,10)$, and $O_{AC}$ represent the pursuer, evader, and the center of the generalized AC respectively. The yellow sphere indicates that $a<1$, the blue surface indicates that $a=1$, and the green surface indicates that $a > 1$ }
\label{Illustration of AC1}
\end{figure}
\section{Proposed Method}
In this section, we extend the 2D AC to the generalized AC in the 3D space. We propose the FACL control algorithm and design a reward function for the agent based on the artificial potential field.

\subsection{Generalized AC in 3D Space}
The ability of agents to engage in capture or evasion depends on the unique input information, specifically the positions and velocities of the agents. We introduce the AC to depict the agent's dominant region and determine their movement strategies based on the geometric relationships between them. 

We extended the AC from 2D space to its generalized form in 3D space, providing the analytical expression of the generalized AC. In Fig.~\ref{Illustration of AC1}, we demonstrate the generalized AC in 3D space under different speed ratio $a = \frac{{{{v_E}}}}{{ {{v_P}} }}$. It can be seen that due to the different values of $a$, the shape of the generalized AC surface also varies. Taking the yellow surface as an example, within the space enclosed by the yellow sphere, $a<1$ is satisfied. Therefore, the evader consistently arrives earlier than the pursuer, thus establishing it as the evader's dominant region. Conversely, the external area represents the pursuer's dominant region.
\begin{figure}[!ht]
\centering
\includegraphics[width=1\columnwidth]{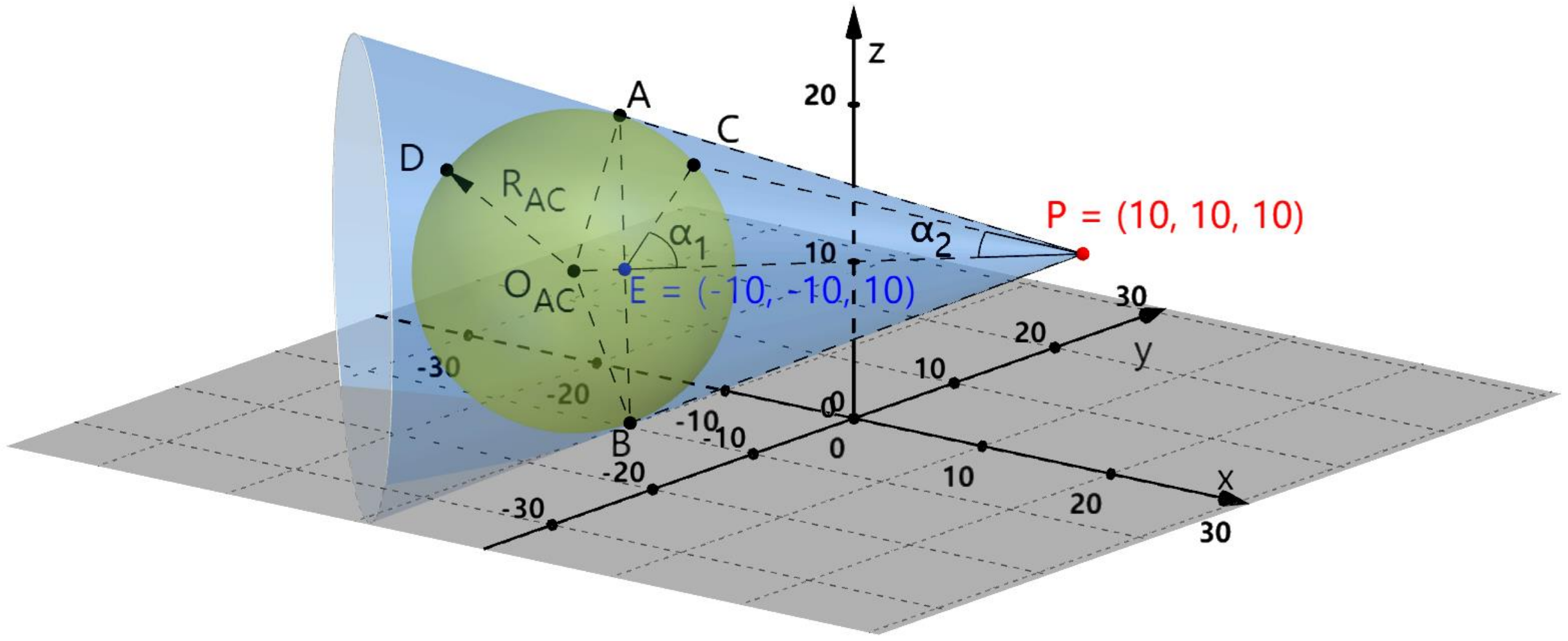}
\caption{Illustration of the generalized AC with $a<1$\label{Illustration of AC}}
\end{figure}

In this paper, we set $a=\frac{{{v_E}}}{{{v_P}}}<1$. The generalized AC composed of the pursuer and the evader is shown in Fig.~\ref{Illustration of AC}. $A$, $B$, $C$ and $D$ are points located on the generalized AC. The lines $PA$ and $PB$ are tangent to sphere $O_{AC}$, that is $PA \bot A{O_{AC}}$ and $PB \bot B{O_{AC}}$, separating the capture and escape regions of the agents. Given the positions of pursuer and evader denoted as $(x_t^P,y_t^P,z_t^P)$ and $(x_t^E,y_t^E,z_t^E)$,  respectively, and the speed ratio $a$ of the agents, we have \eqref{AC-eq}
\begin{equation}
\label{AC-eq}
a = \frac{{{{v_E}}}}{{ {{v_P}} }} = \frac{{\left| {EA} \right|}}{{\left| {PA} \right|}} = \frac{{\left| {EB} \right|}}{{\left| {PB} \right|}} =\frac{{\left| {EC} \right|}}{{\left| {PC} \right|}} < 1,
\end{equation}
where $\left| {EA} \right|$ denotes the Euclidean distance between the evader $E$ and the point $A$. The center and radius of the generalized AC are \eqref{AC-center}
\begin{equation}
\label{AC-center}
\begin{aligned}
    {O_{AC}} &= \left( {\frac{{{x_t^E} - {a^2}{x_t^P}}}{{1 - {a^2}}},\frac{{{y_t^E} - {a^2}{y_t^P}}}{{1 - {a^2}}},\frac{{{z_t^E} - {a^2}{z_t^P}}}{{1 - {a^2}}}} \right)\\
    {R_{AC}} &= \left| {\frac{a}{{1 - {a^2}}}} \right|{\sqrt {{{\left( {{x_t^P} - {x_t^E}} \right)}^2} + {{\left( {{y_t^P} - {y_t^E}} \right)}^2}{\rm{ + }}{{\left( {{z_t^P} - {z_t^E}} \right)}^2}}}.
\end{aligned}
\end{equation}

To ensure that the pursuer always moves towards capturing the evader and that the evader constantly avoids the pursuer's capture, we define the agents' motion strategies. For a random point $C$ on the generalized AC, we have \eqref{AC-C}
\begin{equation}
\label{AC-C}
\frac{{\sin {\alpha _1}}}{{\sin {\alpha _2}}} = \frac{{{v_P}}}{{{v_E}}},
\end{equation}
where $\alpha_1$ is the angle difference between the vector $\overrightarrow {EP}$ and the vector $\overrightarrow {EC}$, and $\alpha_2$ is the angle difference between the vector $\overrightarrow {PE}$ and the vector $\overrightarrow {PC}$. We get \eqref{AC-C1}
\begin{equation}
\label{AC-C1}
{\alpha _2} = {\sin ^{ - 1}}\left( {\frac{{{v_E}}}{{{v_P}}}\sin {\alpha _1}} \right).
\end{equation}
When ${\alpha _1} = \frac{\pi }{2}$, we obtain the optimal solution \eqref{AC-start}
\begin{equation}
\label{AC-start}
\alpha _2^* = {\sin ^{ - 1}}\left( {\frac{{{v_E}}}{{{v_P}}}} \right).
\end{equation}
Therefore, the optimal motion space for both agents lies within the blue conical region shown in Fig.~\ref{Illustration of AC}. The pursuer moves within the envelope defined by the cone, satisfying the condition ${\alpha _2} \le \alpha _2^* = {\sin ^{ - 1}}( {\frac{{{v_E}}}{{{v_P}}}} )$ to efficiently capture the evader. As for the evader, it maneuvers within the remaining cone envelope, satisfying $\alpha_1 \ge \frac{\pi }{2}$.

\subsection{The Fuzzy Actor-Critic Learning Algorithm}
The FACL algorithm, a novel integration of fuzzy logic and actor-critic reinforcement learning (RL) methods, is designed to effectively handle complex decision-making tasks in uncertain environments \cite{Asgharnia2022}. The algorithm comprises two fuzzy logic modules: one functioning as an actor, facilitated by a fuzzy logic controller (FLC), and the other as a critic, employing a fuzzy inference system (FIS). The actor utilizes fuzzy logic to generate action strategies and achieve adaptability in uncertain scenarios. The critic use actor-critic technique to evaluate and guide the training of these strategies.

In this paper, both the actor and critic take first-order Takagi-Sugeno (TS) rules to implement the fuzzy inference system. Assuming the PEG system has $ n $ inputs $\bar x = [{x_1},...,{x_n}] $, the output of the actor is given by \eqref{actor-output}
\begin{equation}
\label{actor-output}
{u_t} = \sum\limits_{l = 1}^L {\Phi _t^l w _t^l}, 
\end{equation}
where $u_t$ is the control signal at time $t$, $L$ is the total number of rules. $ w _t^l = \mathop {\max }\limits_{{u_t},x_i \in \bar{x}} {\mu ^{A_i^l}}({x_i}) $ is the output parameter of the actor for rule $ l $ at time $ t $, where $ {\mu ^{A_i^l}} $ is the membership degree of the fuzzy set $ A_i^l $. The variable $\Phi _t^l$ is the firing strength for rule $ l $ defined as \eqref{firing-strength}
\begin{equation}
\label{firing-strength}
\Phi _t^l = \frac{{\prod\limits_{i = 1}^n {{\mu ^{A_i^l}}({x_i})} }}{{\sum\limits_{l = 1}^L {(\prod\limits_{i = 1}^n {{\mu ^{A_i^l}}({x_i})} )} }}.
\end{equation}
After the actor performs the action, the critic calculates an approximation to ${V_t}$ to evaluate the quality of the action. The output of critic is \eqref{critic-output}
\begin{equation}
\label{critic-output}
{\hat V_t} = \sum\limits_{l = 1}^L {\Phi _t^l\zeta _t^l},
\end{equation}
where $ \zeta _t^l $ is the output parameter of the critic for rule $ l $ at time $ t $. The temporal difference (TD) error $\delta _{t} $ is defined as \eqref{TD-error}
\begin{equation}
\label{TD-error}
\delta _{t} = {r_{t + 1}} + \gamma {\hat V_{t + 1}} - {\hat V_t}.
\end{equation}
The $\delta _{t} $ is then used to update parameters of the actor and critic. To promote exploration of the action space, a random white noise $ {N_0} \sim (0,{\sigma_a ^2}) $ is added to the generated control signal $ {u_t} $. As a result, the output parameter $ w_{t + 1}^l $ of the actor is updated by \eqref{update-actor}
\begin{equation}
\label{update-actor}
w_{t + 1}^l = w_t^l + {\alpha _a}\delta _{t}  \left( {\frac{{{u_t'}  - {u_t}}}{\sigma _a}} \right)\frac{{\partial {u_t}}}{{\partial w_t^l}},
\end{equation}
where ${u_t'}  = {u_t} + {N_0}$, and $ {\alpha _a} $ is learning rate for the actor. The partial derivative $ \frac{{\partial {u_t}}}{{\partial w_t^l}} $ can be calculated by \eqref{partial-derivative-actor}
\begin{equation}
\label{partial-derivative-actor}
\frac{{\partial {u_t}}}{{\partial w_t^l}} = \frac{{\prod\limits_{i = 1}^n {{\mu ^{A_i^l}}({x_i})} }}{{\sum\limits_{l = 1}^L {(\prod\limits_{i = 1}^n {{\mu ^{A_i^l}}({x_i})} )} }} = \Phi _t^l.
\end{equation}
The output parameter $\zeta _{t + 1}^l $ of the critic is updated by \eqref{update-critic}
\begin{equation}
\label{update-critic}
\zeta _{t + 1}^l = \zeta _t^l + {\alpha _c} \delta _{t}  \frac{{\partial {{\hat V}_t}}}{{\partial {\zeta_{t} ^l}}},
\end{equation}
where $ {\alpha _c} $ is learning rate for the critic. The partial derivative $ \frac{{\partial {{\hat V}_t}}}{{\partial \zeta _t^l}} $ can be calculated by \eqref{partial-derivative-critic}
\begin{equation}
\label{partial-derivative-critic}
\frac{{\partial {{\hat V}_t}}}{{\partial \zeta _t^l}} = \frac{{\prod\limits_{i = 1}^n {{\mu ^{A_i^l}}({x_i})} }}{{\sum\limits_{l = 1}^L {(\prod\limits_{i = 1}^n {{\mu ^{A_i^l}}({x_i})} )} }} = \Phi _t^l.
\end{equation}
We set the learning rate $ {\alpha _a} < {\alpha _c} $ so that the actor will converge slower than the critic, preventing instability of the actor.

In this paper, the FACL algorithm is used to learn the optimal strategy for the pursuer and the evader. We define four inputs for each agent. The pursuer's inputs are \eqref{input_P}
\begin{equation}
\label{input_P}
\bar{x}_P = [{d_{PE}},{\delta_{P}}, {d_{PO}}, \delta_{PO}],
\end{equation}
where $ {d_{PE}} $ represents the distance between the pursuer and the evader. The term $ {\delta_{P}}$ denotes the angle difference between the heading of the pursuer ${{{\vec v}_P}(t)}$ and a straight line from the pursuer to the evader ${\overrightarrow {PE} (t)}$. Additionally, ${d_{PO}}$ stands for the distance between the pursuer and the obstacle, and $ \delta_{PO} $ indicates the angle difference between the heading of the pursuer ${{{\vec v}_P}(t)}$ and a straight line from the pursuer to the obstacle ${\overrightarrow {PO} (t)}$. For the evader, the inputs are \eqref{input_E}
\begin{equation}
\label{input_E}
\bar{x}_E = [{d_{PE}},{\delta_{E}}, {d_{EO}}, \delta_{EO}],
\end{equation}
where ${\delta_{E}}$ denotes the angle difference between the heading of the evader ${{{\vec v}_E}(t)}$ and a straight line from the evader to the pursuer ${\overrightarrow {EP} (t)}$. ${d_{EO}}$ is the distance between the evader and the obstacle. The term $ \delta_{EO} $ indicates the angle difference between the heading of the evader ${{{\vec v}_E}(t)}$ and a straight line from the evader to the obstacle ${\overrightarrow {EO} (t)}$.

\subsection{Reward Function Based on Artificial Potential Field}

As we all know, the reward function plays a crucial role in RL, as it significantly influences both the convergence speed and overall performance of the RL algorithm. In this paper, we design a reward function based on the idea of the artificial potential field. The obstacles exert repulsive force on the pursuer, whereas the evader exerts attraction on the pursuer. We model the repulsion exerted by obstacles on the pursuer using an exponential function. The reward function for repulsion can be designed as \eqref{Rpo}
\begin{equation}
\label{Rpo}
r_{PO} = 1 - exp(-\alpha_r * \Delta d_{PO}),
\end{equation}
where $\Delta {d_{PO}} = {d_{PO}}(t + 1) - {d_{PO}}(t)$, and $\alpha_{r}$ is the repulsion coefficient controlling the strength of the repulsive force. The evader has the same form of reward function for obstacle. The reward function for attraction can be formulated as
\begin{equation}
\label{RPE}
r_{PE} = exp({ - {\beta _a}\Delta {d_{PE}}}) - 1,
\end{equation}
where $\Delta {d_{PE}} = {d_{PE}}(t + 1) - {d_{PE}}(t)$, and $\beta_{a}$ is the attraction coefficient governing the strength of the attractive force. For the evader, \eqref{RPE} is inverted. Incorporate an additional term to the reward function that considers whether the pursuer successfully captures the evader. This can be achieved by adding a success-based reward term
\begin{equation}
\label{success-based reward}
    r_{s} = \gamma_s * g_{s},
\end{equation}
where $\gamma_s$ is the coefficient for the success-based reward, and $g_{s}$ is an indicator function that $g_{s} = 1$ when the pursuer captures the evader, and $g_{s} = 0$ otherwise. This component motivates the pursuer to complete its objective successfully. For the evader, \eqref{success-based reward} is inverted. We formulate a comprehensive reward function using a weighted balance between these forces
\begin{equation}
\label{comprehensive reward}
r_{total }= w_r*r_{PO} + w_a*r_{PE} + r_{s},
\end{equation}
where $w_r$ and $w_a$ represent the weights for balancing the influence of repulsion and attraction. By adjusting these weights, we can control the overall behavior of the reward function and efficiently complete the PEG.

Throughout the learning process, each agent selects an action based on the current state. After taking the action and receiving the corresponding reward, the actors and critics perform parameter updates by \eqref{update-actor} and \eqref{update-critic}. The learning process then continues until the final state is reached, where the evader is captured by the pursuer, that is $d_{PE}(t_f) \le d_s$, or the maximum learning time is reached the designated threshold.

\section{Simulation and Result}
We evaluated the learning performance of the FACL algorithm in a PEG scenario. 
We define a 3D environment for the PEG, where agents move in a space with a size of $35{\rm{m}} \times 35{\rm{m}} \times 20{\rm{m}} $ at maximum speeds of $v_P = 1.1{\rm{m/s}}$ and $v_E = 1{\rm{m/s}}$, respectively. According to \eqref{comprehensive reward}, the parameters of the reward functions are set as ${\alpha _r} = 10,{\beta _a} = 5,{\gamma _s} = 20, {w_r} = 5,{w_a} = 10$. The number of episodes is set to 200, and the number of plays in each game is 500. The sample time is $ 0.1\rm{s}$. The discount factor is $ \gamma  = 0.95 $, the random white noise is $ {N_0} \sim (0,{0.01}) $, and the learning rates are $ {\alpha _a}=0.001, {\alpha _c} = 0.05 $. The game terminates when the pursuer captures the evader, that is ${d_{PE}} \le {d_s} = 1\rm{m}$ or when the time exceeds 100 seconds. To reduce the computational load, we chose the triangular membership function. Specifically, the FACL has 5 triangular membership functions for each input. The distance inputs are in the interval of $ [0, 35] $, and the angle inputs are in the interval of $[-\pi, \pi]$. Consequently, we have $5 \times 5 \times 5 \times 5 = 625$ rules. In the 3D environment, we select the output of the agent controller from a set of tuples $\left\{ {(\alpha,\theta)|\alpha,\theta \in [ - \frac{\pi }{4},\frac{\pi }{4}]} \right\}$.

\begin{figure}[!ht]
\centering
\includegraphics[width=0.9\columnwidth]{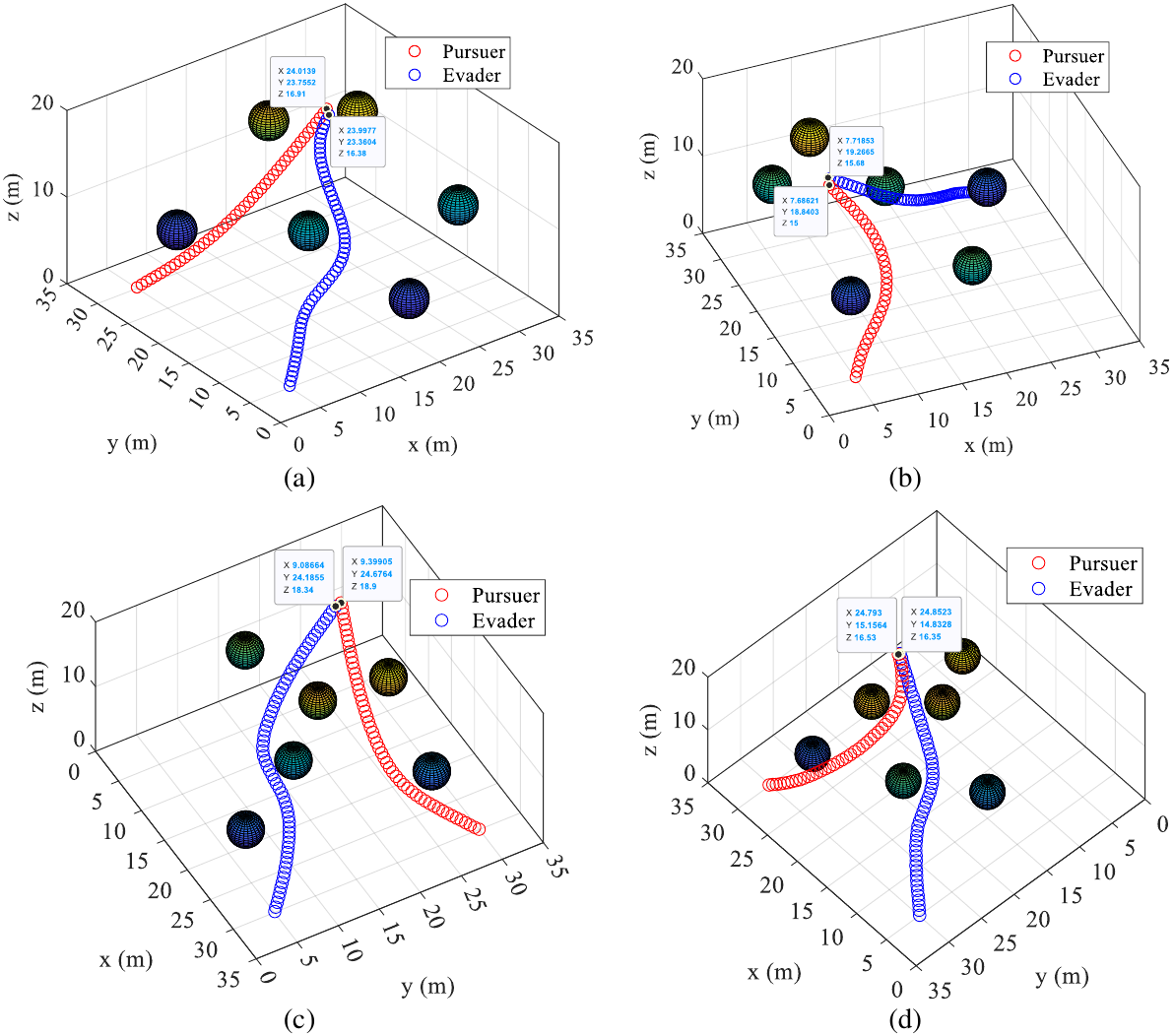}
\caption{Trajectories of agents in the 3D environment \label{Trajectories-3d}}
\end{figure}

\begin{figure}[!ht]
\centering
\includegraphics[width=0.75\columnwidth]{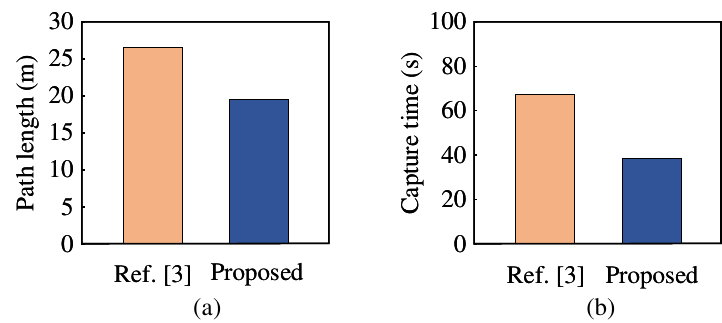}
\caption{The path length and capture time of two algorithms  \label{Histogram}}
\end{figure}

We have set up four different scenarios. The positions of pursuers are $[5,30,0]$, $[5,5,0]$, $[30,30,0]$, and $[30,30,0]$, while the corresponding positions of evaders are $[5,5,0]$, $[30,30,0]$, $[30,5,0]$, and $[5,30,0]$. The obstacles are randomly placed and are represented by spheres with a radius of $1{\rm{m}}$. The trajectories of pursuers and evaders are illustrated in Fig.~\ref{Trajectories-3d}. We can observe that in the 3D environment, the pursuer successfully captured the evader, without any collisions occurring. After the PEG ends, the distances between the pursuer and the evader are 0.66, 0.80, 0.81, and 0.38, respectively, all of which are less than the designed distance $d_s=1\rm{m}$. In \cite{Al-Mahbashi2020}, the passive 3D PEG with interception was studied. Compared to the algorithm in \cite{Al-Mahbashi2020}, the algorithm proposed in this paper enables the agent to engage in a more proactive pursuit of the evader, exhibiting greater flexibility in maneuvering and superior adaptability. We conducted 20 repeated tests on the algorithm in \cite{Al-Mahbashi2020} and the proposed algorithm, and calculated the average of the results. As illustrated in Fig.~\ref{Histogram}, a comparison was made between the two algorithms based on the parameters of the path length and capture time. Our proposed algorithm outperforms the method in \cite{Al-Mahbashi2020} in both aspects.

\section{Conclusions}
This paper investigates the PEG in a 3D space. We have extended the 2D AC framework to a 3D space and presented its analytical form. By analyzing the capture condition, we delineate the optimal motion space for the agents. We designed a FACL algorithm for the agents and developed a reward function based on the concept of artificial potential field to achieve efficient PEG. The effectiveness of our algorithm has been verified through simulation result. Future research will concentrate on PEG involving multiple agents in a 3D space, as well as multi-objective RL algorithms to address PEG problems with multiple optimization objectives.


\end{document}